# A hands-on laboratory and computational experience for nanoscale materials, devices and systems education for electronics, spintronics and optoelectronics

Hassan Raza[1], Tehseen Z. Raza[2]

[1] Department of Electrical and Computer Engineering, University of Iowa, Iowa City, Iowa 52242, USA
nstnrg@gmail.com
[2] Department of Physics and Astronomy, University of Iowa, Iowa City, Iowa 52242, USA
tehseen-raza@uiowa.edu

Abstract: To enhance the undergraduate and graduate engineering education for nanoscale materials, devices and systems, we report a multi-disciplinary course based on the integration of theory, hands-on laboratory and hands-on computation into a single curriculum. The hands-on laboratory modules span various dimensionalities of nanomaterials as well as applications in logic, memory, and energy harvesting. In the hands-on computational exercises, students simulate the material and the device characteristics, and in some cases, design the experimental process flow to fabricate and characterize the devices and systems. Such a course not only grooms the students for multi-disciplinary collaborative activities in nanoscience and nanoengineering, but also prepares them well for future academic or industrial pursuit in this area.

## 1. INTRODUCTION

Smart materials, multifunctional devices, self-healing systems and novel applications are emerging due to exotic quantum phenomenon at the nanoscale. Nanotechnology [1] is expected to impact far-reaching scientific and technological areas, such as medicine, electronics, biomaterials, energy harvesting, communications, manufacturing, transportation, space exploration - to name a few [2,3].

With the devices shrinking in size to nanoscale, not only the packing density and speed are improving, but there is also a need of improved functionality and sensitivity. Novel physical phenomena, like the spin of an electron instead of its charge, may become the leading role player in the future devices. Nanotechnology is also expected to provide a solution to extend the Moore's Law in this century by laying down the foundations of novel logic and memory devices. The next generation of sustainable energy technologies will be based on transformational new materials that convert energy efficiently among photons, electrons, and chemical energy. In this context, an engineer must be equipped with all the necessary theoretical, computational and experimental tools to understand the material properties for the device design and system behavior [4-17].

The top-down approach for device manufacturing has been successfully used at the macro and the micron scale. This paradigm is reaching its limits as we approach the nanometer dimensions due to lack of atomic control. At the nanoscale, the bottom-up approach [18] thus becomes a compulsion not only in theoretical design by using atomistic models, but also in material synthesis and device fabrication. A classical example is the molecular self-assembly where nature does the work with properly-engineered molecules. Within this multi- and inter-disciplinary bottom-up approach, the physical, chemical, biological sciences and engineering have all arrived at the nanoscale. The vastly different phenomena and challenges in these areas can thus be cast into atomic scale details in a coherent yet comprehensive manner, thus providing a common platform for all these seemingly-different fields.

| MATERIALS | DEVICES | SYSTEMS |
|---|---|---|
| • <u>Zero-dimensional</u><br>  - Molecules<br>  - Quantum Dots<br>  - Nanocrystals<br><br>• <u>One-dimensional</u><br>  - Carbon nanotubes<br>  - Silicon nanowires<br>  - Nanoribbons<br><br>• <u>Two-dimensional</u><br>  - Graphene<br>  - Surfaces<br><br>• <u>Three-dimensional</u><br>  - Heterostructures<br>  - Magnetic Tunnel junctions | • <u>Information Processing</u><br><br>  - Electronics<br>  - Spintronics<br>  - Photonics<br>  - Sensors<br>  - Biomedical devices<br><br>• <u>Energy</u><br><br>  - Solar Cells<br>  - Thermoelectrics<br>  - Batteries | • <u>Information Processing</u><br><br>  - Nano electro-mechanical systems (NEMS)<br>  - Charged coupled devices (CCDs)<br><br>• <u>Energy</u><br>  - Integrating thermoelectrics and solar cells |

Figure 1: The curriculum comprises of understanding nanotechnology using the bottom-up approach, starting from the nanomaterials to devices and finally systems, from the theoretical and experimental aspects complemented by the state-of-the-art in computation.

## 2. TEACHING PHILOSOPHY:

To develop an engineering approach towards nanotechnology, we report the development of a multi-disciplinary course with a hands-on laboratory [13,19,20] and computational experience for the senior undergraduate and first-year graduate students. It introduces students to the complete nanodevice design space from an integrated approach of theoretical and computational understanding, complemented with an experimental experience. The important aspect is to explore the depth and breadth of this field starting from nanomaterials (zero-dimensional, one-dimensional, two-dimensional, three-dimensional) to nanofabrication, simulation and analysis of the devices and then finally system integration as shown in Fig. 1 [21]. Theory and experiments are designed to go hand-in-hand, both complemented with various computational modules. This is to give a broad, yet detailed, introduction to the field of nanotechnology from an engineering perspective, both at device [electronics, spintronics, optoelectronics] and system [CCD (charged couple device), NEMS (nano-electromechanical system)] level. The CCD discussion specifically brings the system-level aspect in this curriculum, where array-connected nanoscale transistors can perform useful signal processing, memory, sensing and imaging tasks. Similarly, carbon nanotube based NEMS are analyzed for potential applications as oscillators and sensors.

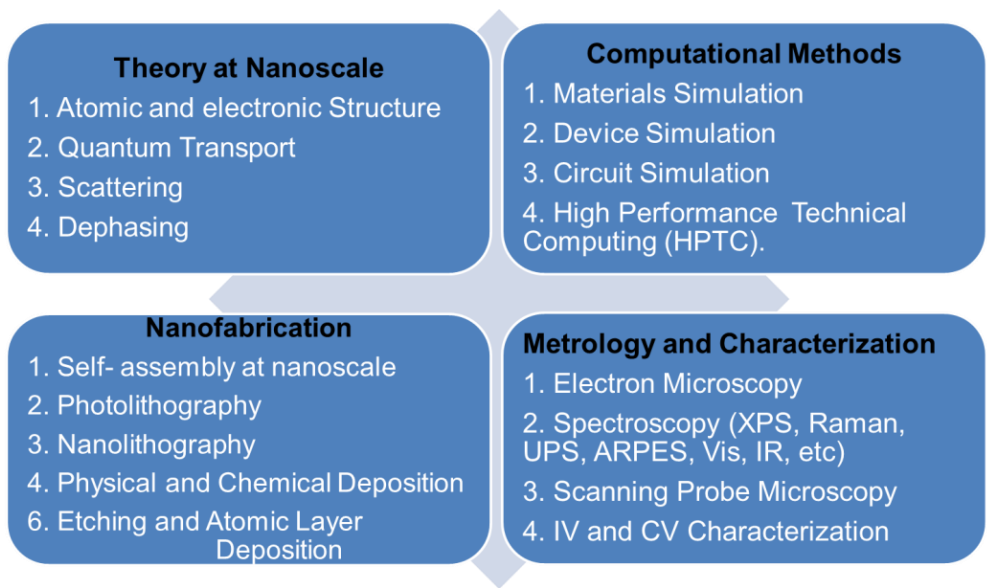

Figure 2: Design and analysis of nanoscale materials, devices and systems based on the integration of theory, computation and laboratory modules (nanofabrication and characterization). The important topics to be addressed in each section are highlighted.

The theoretical thrust of understanding the materials, devices, and system is highlighted in Fig. 2, starting from the material properties to developing an understanding of the quantum transport in electronics, spintronics, photonics, sensors and energy harvesting devices. Various dephasing mechanisms are also discussed to understand non-ideal behavior. The process flow of device fabrication is then discussed followed with experimental demonstrations and hands-on laboratory work of lithography and device fabrication. Theory of various device characterization techniques is taught in the class, complemented with laboratory tours and various demonstrations to see how the characterization and spectroscopy equipment works. Finally, various systems are analyzed by building upon the device characteristics. The hand-on laboratory for the students is devised in such a way as to give them exposure to basic device fabrication scheme. After the device fabrication, another laboratory module requires them to perform the device characterization.

## 2.1 THEORETICAL MODELING

The bottom-up approach to nanotechnology provides a common framework and test bed to integrate various concepts at the atomic scale. It is a natural and intuitive way of understanding devices where the building-blocks are atoms. The theoretical thrust of the course, complemented with hands-on computational modules, converges to focusing on the following main aspects,

- Electronic structure theory to determine the properties of the nanomaterials (a solution of the Schrödinger equation).

- Quantum transport to study the carrier flow at the nanoscale by solving the Schrödinger equation with open boundary conditions in the form of contacts.

- Statistical mechanics for carrier (fermion, boson) statistics

- Scattering processes to understand the non-ideal effects.

To understand the material properties, we teach the electronic structure at various levels of detail and sophistication. Starting from the effective mass model and empirical tight-binding theory, we cover the semi-empirical tight-binding methods (e.g. extended Hückel theory, etc) and ultimately *ab initio* methods (e.g. Hartree-Fock theory and density functional theory, etc.). Exposing students to varying levels of detail helps to give them the big picture in nanotechnology design and integration. Indeed, all these methods have different levels of sophistication and depending upon the resource availability and the accuracy requirements, one would decide on a trade-off between them.

Once the electronic structure of a material is well-established, the quantum transport is the next step, for which, we use non-equilibrium Green's function (NEGF) formalism [22,23], which is a solution to the Schrödinger's equation with open boundary conditions in the form of contacts within the mean-field picture. NEGF has become a standard method of choice that provides a unified approach to study the non-equilibrium phenomena for a wide range of materials and systems from molecular to ballistic to diffusive transport. It can inherently be applied to understand the charge and heat transport in electronic, spintronic, photovoltaic and thermoelectric devices. Finally the non-ideal effects in the form of scattering and dephasing are included within the mean-field self-consistent Born approximation (SCBA) [23-26]. Since the electronic structure, NEGF and SCBA equations depend upon each other, the students are taught how to solve these equations self-consistently.

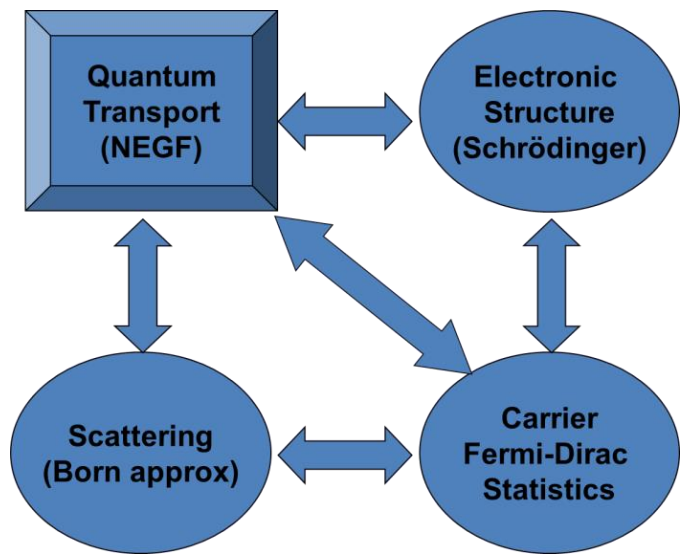

Figure 3: Non-equilibrium devices and systems at the nanoscale. For logic, memory and energy-harvesting applications, one has to understand atomic and nanoscale devices and systems under highly non-equilibrium conditions. Such analysis may involve a self-consistent picture of concepts and equations across disciplines of quantum mechanics, statistical mechanics, open systems and scattering.

## 2.2 COMPUTATIONAL SIMULATIONS

At the heart of the reported curriculum are the MATLAB-based material and device simulations for understanding nanoscale materials, devices, and finally the system design. Students learn about the effect of dimensionality on the electronic-structure as well as charge distribution on various atoms. They not only simulate the practice examples, but create new nanostructures and even develop their own computational modules to add further functionality. Once the equilibrium material characteristics are well-understood, the next step is to add contacts to make devices and calculate the transport properties using NEGF, followed by connecting multiple devices to study system-level simulations, e.g. charge coupled devices, memories, NEMS, etc. As homework assignments, students develop NEGF based transport models for silicon [26-28] and carbon based nanostructures [29-39], as well as Fe-MgO-Fe magnetic tunnel junctions [40,41]. The following device calculations are few practice examples taught to them in class,

- Bandstructure calculations for various materials like graphene, nanoribbons, etc.
- I-V characteristics TB-NEGF calculations through PN junction diode, Resonant Tunneling Diode and Field Effect Transistor
- Flash memory device characteristics
- MTJ device characteristics
- Thermoelectric device characteristics

Specific examples discussed in the course for Sections 2.1 and 2.2 are reported in Ref. [42].

**2.3 FABRICATION, CHARACTERIZATION, MICROSCOPY, AND SPECTROSCOPY**

The experimental equipment for fabrication, characterization, microscopy and spectroscopy is explained in the class lectures. We start with the standard photolithography methods. After discussing their spatial limitations, we introduce the students to the state of the art nanofabrication methods, e.g. electron-beam lithography and nano-imprint lithography. Furthermore, physical and chemical vapor deposition are discussed with the applications of these methods to fabricating nanocrystals, quantum dots, carbon nanotubes, graphene, etc, in a bottom up fashion. We further discuss molecular self-assembly to understand the bottom-up approach towards nanotechnology. Novel deposition methods like atomic-layer deposition (ALD) are discussed, which has become a method of choice for the gate stack in the CMOS (complementary metal-oxide-semiconductor) technology.

Various imaging (optical, electron) methods are then discussed, which include but are not limited to confocal microscopy, scanning electron microscopy (SEM), transmission electron microscopy (TEM) and scanning transmission electron microscopy (STEM). Furthermore, we introduce scanning probe microscopes and their diverse capabilities and variations in the form of scanning tunneling microscope (STM) and atomic force microscope (AFM).

Finally, we focus on the spectroscopic techniques, which include scanning probe spectroscopy (SPS), scanning tunneling spectroscopy (STS), X-ray photoemission spectroscopy (XPS), ultra-violet photoemission spectroscopy (UPS), angle-resolved photoemission spectroscopy (ARPES), Raman Spectroscopy, FTIR (Fourier transform Infrared) spectroscopy, etc. These discussions complement the various simulation exercises and also, in some cases, hands-on laboratory modules (see next section).

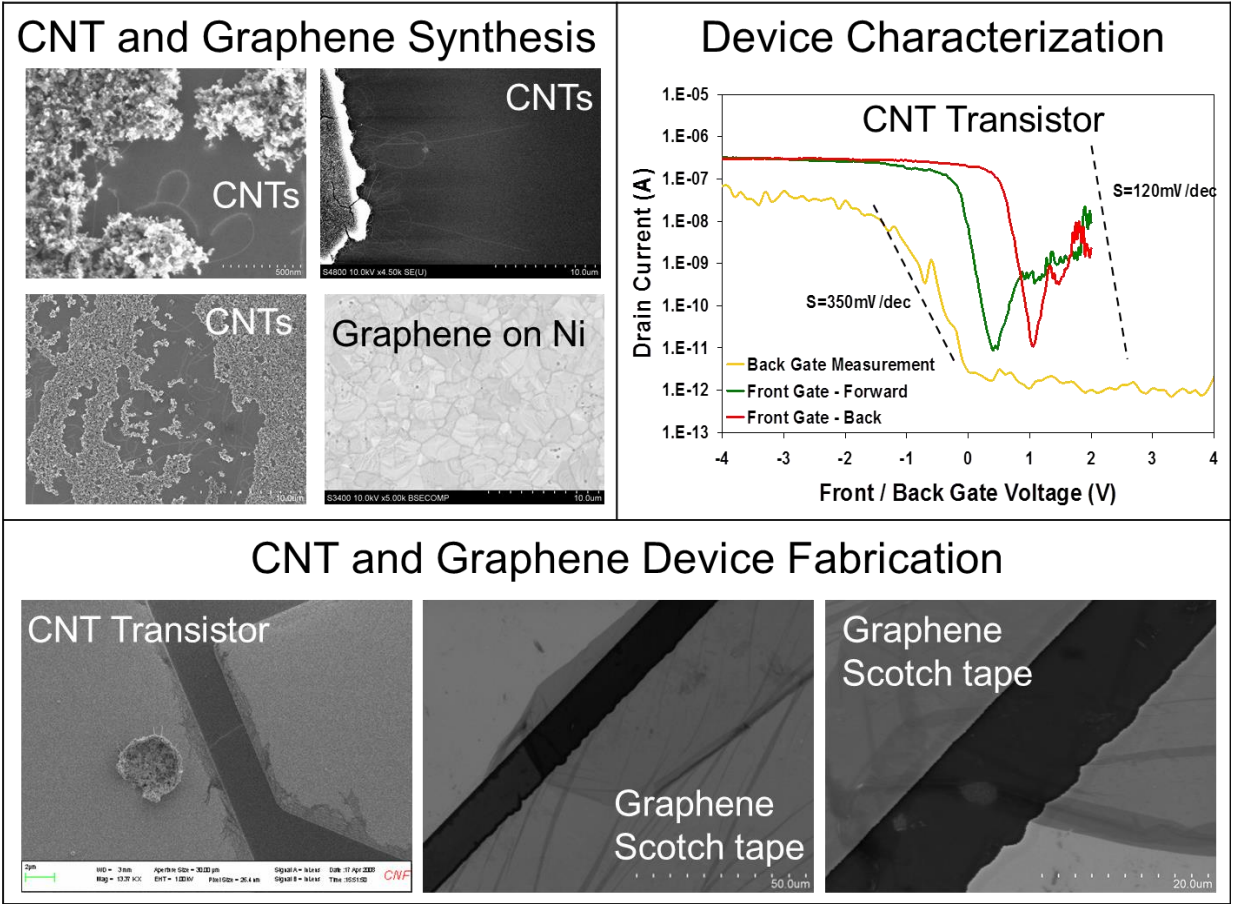

Figure 4: A few examples of material and device characterization. SEM viewgraphs of carbon nanotube and graphene samples synthesized using CVD method. The device characterization of a CNT transistor is shown with a better inverse sub threshold slope (S) using the top gate. Under device fabrication, CNT transistor and Scotch tape exfoliated graphene transistor are shown.

## 2.4 HANDS-ON LABORATORY EXPERIENCE

Another practical aspect of this curriculum is the hands-on laboratory experience by mostly using home-made equipment. These laboratory modules are designed to introduce the basic concepts in device fabrication, the techniques and the tools that are central to the rapidly developing field of nanoscience and nanotechnology, and to some of the important scientific and technological innovations that are now emerging, or are expected to emerge, from this field. The home-made equipment (like a CVD furnace) enlightens and inspires students to engineer their own laboratory equipment when required. The following laboratory modules were designed as a part of the course. The fabrication process flows and characterization details for these laboratory modules are available in Ref. [42].

**i) Introductory lab module:** Techniques learned in this laboratory exercise are photolithography, physical vapor deposition (evaporation), metal annealing and optical microscopy imaging, where students incorporate gold nanocrystals in a gate stack to form a flash memory.

**ii) Carbon Nanotube / Graphene lab module:** A carbon nanoelectronics laboratory module involves experiments to grow carbon nanotubes (CNTs) using homemade chemical vapor deposition (CVD) equipment. CNTs were grown in methane ambient at 950°C at atmospheric pressure by using Fe based nanoparticle catalyst [28-31]. These two samples along with a fullerene sample is characterized using Raman spectroscopy and imaged using SEM. For the CNT samples, we look for the radial breathing modes to determine the CNT diameter, which was found to be in 1-2 nm range [28-31]. Statistically, one-third CNTs are semi-metallic, whereas the rest are semi-conducting. These semiconducting CNTs are important for device applications. We report the transport properties of one such semiconducting CNT with back (bottom) gate and front (top) gate. Additionally, we deposit graphene between contacts using a Scotch tape method as shown in Fig. 4.

**iii) Energy lab module:** This laboratory module involves experiments related to dye-sensitized solar cells by using commercially available kits [43]. The students make solar cells using commonly available fruit juices and study their output characteristics.

Additional details about the recipes followed in the hands-on laboratory modules are provided in Ref. [42]. Students developed basic fabrication skills in the introductory laboratory module and then followed the recipes to fabricate their own devices from scratch.

## 3. STUDENT SURVEY RATINGS:

This course has been offered in Fall 2011 at the University of Iowa as an elective course for junior/senior undergraduate and first-year graduate students in Engineering, Physics and Chemistry. Eleven students registered for the course, which is a nominal class size for a course at this level in a small department. Amongst them, two students were female. Eight students were from Electrical Engineering and three were from Physics department. Junior/Senior/Graduate standings were 1/4/6, respectively. An elementary device course at junior level was the pre-requisite for the course. The student feedback statistics and distributions are shown in Table I and Fig. 5, respectively. Additional comments are provided below,

- I really enjoyed the laboratory modules; they were a nice added learning experience to a lot of the things we were talking about in class.

- The material in this class was really interesting even with the complexity of the subject matter. It has been covered many different and versatile areas related to nanoscale devices and systems.
- The actual strength of the course was in the first half of the course where the theoretical computational models were discussed.
- The class notes are excellent and will come in handy many years from now.
- This course gave a great introduction to nanoscale devices.

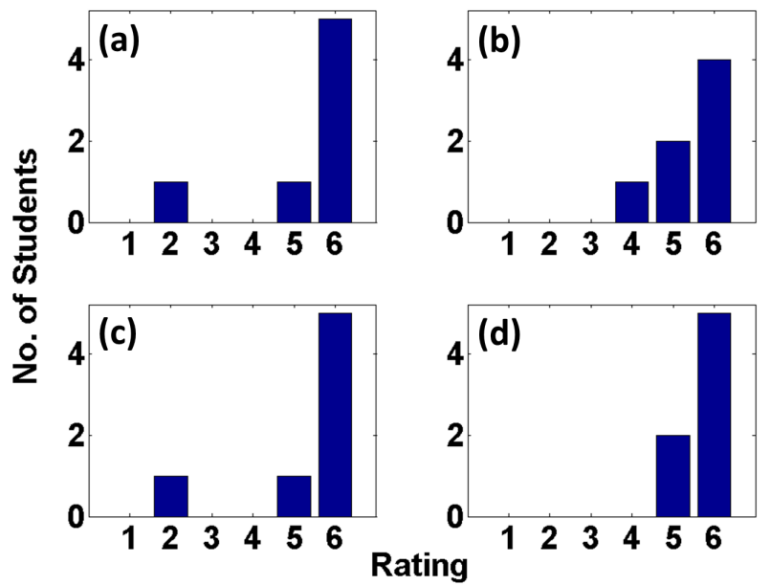

Figure 5: Student feedback distribution. (a) Effectiveness of instruction, (b) Acquiring a basic understanding of nanoengineering, (c) Organization of the course, (d) Contribution of the instructional methods to learning.

Table I: Student feedback statistics.

|  | Fall 2011 | |
| --- | --- | --- |
|  | Mean | Median |
| Effectiveness of Instruction | 5.29 | 6.00 |
| Acquiring a basic understanding of Nanotechnology | 5.43 | 6.00 |
| Organization of the course | 5.29 | 6.00 |
| Contribution of the instructional methods to learning | 5.71 | 6.00 |

6 = Strongly Agree

5 = Moderately Agree

4 = Slightly Agree

3 = Slightly Disagree

2 = Moderately Disagree

1 = Strongly Disagree

0 = Skipped

## 4. FUTURE DIRECTIONS:

We proposed the following future developments for this course, which would provide more depth to the topics already covered.

**i) Spintronics experimental module:** We plan to include a laboratory module on Spintronics devices by fabricating a $Co/Al_2O_3/Co$ magnetic tunnel junction memory device and performing magneto-electrical characterization.

**ii) Home-made AFM Laboratory module:** AFM is one of the foremost tools of nanotechnology for imaging, measuring, and manipulating the matter at the nanoscale. Understanding the basic forces that govern the imaging of nanomaterials and manipulating them will increase the physical understanding of students. We are in the process of developing a home-made AFM, which will be used in a hands-on laboratory module. Students will learn all the details of AFM working principles and will use it to image nanomaterials and nanodevices. Such a device is also a working proof of a physical concept incorporated into a practical device, which is a huge learning experience itself.

**iii) High Performance Technical Computing (HPTC) Module:** We further plan to include HPTC in the course, where the speed advantages brought by parallel processing using multiple cores would be demonstrated to the students. The ability to create programs which can run on multiple cores, nodes or computers can speed up the computation tremendously. This part will include discussion related to hyper threading and message passing interfacing in C and Matlab. Furthermore, students will also learn to make a computing cluster using Linux.

## 5. CONCLUSIONS:

We have successfully developed a nanoengineering devices and systems course that covers the depth and breadth of device design from theoretical, experimental and computational aspects. The broader aspect of the integrated curriculum will produce engineers with the right skills to understand nanotechnology and nanoengineeirng. Generally, lack of faculty (mostly in small departments like the Department of Electrical and Computer Engineering at the University of Iowa) has been the main impeding factor for a comprehensive engineering nanotechnology curriculum at the undergraduate and graduate level. Training and education provided through the reported multi-disciplinary curriculum will produce well-trained engineers in nanotechnology, who will be ready to solve the technological and societal problems of today and tomorrow. These students will be equipped with the theoretical knowledge base and the right skills to propose, fabricate and then analyze the nanoscale devices. We anticipate that with the skills acquired, students will be well-equipped for not only multi-disciplinary research collaborations but also for the industry and beyond.

**Acknowledgements:**

We thank A. Umair for helpful discussions about the laboratory modules. We are also grateful to A. Umair, A. Mohsin, T.-H. Hou, J. Baltrusaitis, K. Hutchinson and F. Wang for SEM, CNT, and graphene sample help. We would like to acknowledge the Provost office at the University of Iowa for financial support.

**REFERENCES:**

[1] N. Taniguchi, "On the Basic Concept of 'Nano-Technology'", Poc. Intl. Conf. Prod. Eng. Tokyo, Part II, Japan Society of Precision Engineering (1974).

[2] M. C. Roco, "Nanotechnology Research Directions for Societal Needs in 2020", WWCS, December 1 (2010).

[3] M. C. Roco, "National Nanotechnology Initiative - Past, Present, Future", Handbook on Nanoscience, Engineering and Technology, (Second Edition, Taylor and Francis, 2007).

[4] M. C. Roco, "Nanoscale Science and Engineering: Unifying and Transforming Tools", AIChE Journal **50**, 890 (2004).

[5] M. C. Roco, "Converging Science and Technology at the Nanoscale: Opportunities for Education and Training", Nature Biotechnology **21**, 1247 (2003).

[6] N. Healy, "Why Nano Education?", Journal of Nano Education **1**, 6 (2009).

[7] S. J. Fonash, "Education and Training of the Nanotechnology Workforce", Journal of Nanoparticle Research **3**, 79 (2001).

[8] W. Zheng, H.-R. Shih, K. Lozano, J.-S. Pei, K. Kiefer, and X. Ma, "A Practical Approach to Integrating Nanotechnology Education and Research into Civil Engineering Undergraduate Curriculum", J. Nano. Educ. **1**, 22-33 (2009).

[9] K. Winkelmann, "Practical Aspects of Creating an Interdisciplinary Nanotechnology Laboratory Course for Freshmen", J. Nano. Educ. **1**, 34-41 (2009).

[10] M. Al-Haik, C. Luhrs, Z. Leseman, and M. R. Taha, "Introducing Nanotechnology to Mechanical and Civil Engineering Students Through Materials Science Courses", J. Nano Educ. **2**, 13-26 (2010).

[11] T. R. Tretter, M. G. Jones, and M. Falvo, "Impact of Introductory Nanoscience Course on College Freshmen's Conceptions of Spatial Scale", J. Nano Educ. **2**, 53-66 (2010).

[12] M. Uddin, A. R. Chowdhury, "Integration of Nanotechnology into the Undergraduate Engineering Curriculum", International Conference on Engineering Education August 6 – 10, 2001 Oslo, Norway.

[13] D. V. Russo, M. J. Burek, R. M. Iutzi, J. A Mracek, and T. Hesjedal, "Development of an Electronic Nose Sensing Platform for Undergraduate Education in Nanotechnology", Eur. J. Phys. **32**, 675 (2011).

[14] A. Barrañón, A. Juanico, "Major Issues in Designing an Undergraduate Program in Nanotechnology: the Mexican Case", WSEAS Transactions on Mathematics **9**, 264 (2010).

[15] P. Goodhew, "Education Moves to a New Scale", Nanotoday **1**, 40 (2006).

[16] K. P. Chong, "Nanoscience and Engineering in Mechanics and Materials", Journal of Physics and Chemistry of Solids **65**, 1501 (2004).

[17] http://nanohub.org/education/nanocurriculum, when accessed.

[18] M. Vaidyanathan, "Electronics From the Bottom Up: Strategies for Teaching Nanoelectronics at the Undergraduate Level", IEEE Transactions on Education **54**, 77 (2011).

[19] J. D. Adams, B. S. Rogers, L. J. Leifer, "Microtechnology, Nanotechnology, and the Scanning-Probe Microscope: an Innovative Course", IEEE Transactions on Education **1**, 51 (2004).

[20] D. W. Lehmpuh, "Incorporating Scanning Probe Microscopy into the Undergraduate Chemistry Curriculum", J. Chem. Educ. **80**, 478 (2003).

[21] For further details, see https://sites.google.com/site/nstnrg/home/teaching/ui195fa11

[22] S. Datta, Electronic Transport in Mesoscopic Systems (Cambridge University Press, Cambridge, UK, 1997).

[23] S. Datta, Quantum Transport: Atom to Transistor (Cambridge University Press, Cambridge, UK, 2005).

[24] H. Raza, E. C. Kan, "An Atomistic Quantum Transport Solver with Dephasing for Field-Effect Transistors", J. Comp. Elec. **7**, 423 (2008).

[25] H. Raza, "A Theoretical Model for Single Molecule Incoherent Scanning Tunneling Spectroscopy", J. Phys.: Condens. Matter **20**, 445004 (2008).

[26] H. Raza, K. H. Bevan and D. Kienle, "Incoherent Transport Through Molecules on Silicon in the Vicinity of a Dangling Bond", Phys. Rev. B **77**, 035432 (2008).

[27] H. Raza, "Theoretical Study of Isolated Dangling Bonds, Dangling Bond Wires and Dangling Bond Clusters on H:Si(100)-(2×1) surface", Phys. Rev. B **76**, 045308 (2007).

[28] H. Raza and E. C. Kan, "Electrical Transport in a Two-dimensional Electron and Hole Gas on a Si(001)-(2×1) surface", Phys. Rev. B **78**, 193401 (2008).

[29] R. Saito, G. Dresselhaus, and M. S. Dresselhaus, Physical Properties of Carbon Nanotubes, (Imperial College Press, London, UK, 1998).

[30] M. S. Dresselhaus; G. Dresselhaus, P. Avouris, Carbon Nanotubes: Synthesis, Structure, Properties and Applications, (Springer, Berlin Heidelberg New York, 2001).

[31] M. S. Dresselhaus; G. Dresselhaus, P. C. Eklund, Science of Fullerenes and Carbon Nanotubes: Their Properties and Applications, (Academic Press, San Diego, California, 1996)

[32] A. Jorio, M. S. Dresselhausa; G. Dresselhaus, Carbon Nanotubes: Advanced Topics in the Synthesis, Structure, Properties and Applications, (Springer, Berlin Heidelberg New York, 2008).

[33] H. Raza, "Passivation and Edge Effects in Armchair Graphene Nanoribbons", Phys. Rev. B **84**, 165425 (2011).

[34] H. Raza and E. C. Kan, "Armchair Graphene Nanoribbons: Electronic Structure and Electric-field Modulation", Phys. Rev. B **77**, 245434 (2008).

[35] H. Raza and E. C. Kan, "Field Modulation in Bilayer Graphene Band Structure", J. Phys.: Condens. Matter **21**, 102202 (2009) - Fast Track Communication.

[36] H. Raza, "Zigzag Graphene Nanoribbons: Bandgap and Midgap State Modulation", J. Phys.: Condens. Matter **23**, 382203 (2011) – Fast Track Communication.

[37] H. Raza, Graphene Nanoelectronics: Metrology, Synthesis, Properties and Applications, (Springer, Berlin Heidelberg New York, 2011)

[38] A. K. Geim, "Graphene: Status and Prospects", Science **324**, 1530 (2009).

[39] H. Raza and E. C. Kan, "An Extended Hückel Theory Based Atomistic Model for Graphene Nanoelectronics", J. Comp. Elec. **7**, 372 (2008).

[40] T. Z. Raza, H. Raza, "Independent-band Tight-binding Parameters for Fe-MgO-Fe Magnetic Heterostructures", IEEE Trans. Nanotechnol. **10**, 237 (2011).

[41] T. Z. Raza, J. I. Cerda, H. Raza, "Three-dimensional Extended Hückel Theory-Nonequilibrium Green's Function Spin Polarized Transport Model for Fe/MgO/Fe Heterostructures", J. Appl. Phys. **109**, 023705 (2011).

[42] H. Raza, Nanoelectronics Fundamentals, Materials, Devices and Applications, in preparation.

[43] http://ice.chem.wisc.edu/, when accessed.